\documentclass[aps,prd,reprint,twocolumn,superscriptaddress,showpacs]{revtex4-1}

\usepackage{graphicx}
\usepackage{mathrsfs}
\usepackage{bm}
\usepackage{amsmath}
\usepackage{dcolumn}
\usepackage{epstopdf}
\usepackage{dsfont}
\usepackage{amssymb}
\usepackage{tabularx}
\usepackage{array}
\usepackage{float}
\usepackage{color}
\usepackage{epstopdf}
\usepackage{mathrsfs}
\usepackage[colorlinks, linkcolor=blue,anchorcolor=blue,citecolor=blue,urlcolor=blue]{hyperref}

\begin{document}
	
\title{Valley-dependent properties of monolayer MoSi$_{2}$N$_{4}$, WSi$_{2}$N$_{4}$ and MoSi$_{2}$As$_{4}$}

\author{Si Li}
\affiliation{School of Physics and Electronics, Hunan Normal University, Changsha, Hunan 410081, China}
\affiliation{Research Laboratory for Quantum Materials, Singapore University of Technology and Design, Singapore 487372, Singapore}

\author{Weikang Wu}
\affiliation{Research Laboratory for Quantum Materials, Singapore University of Technology and Design, Singapore 487372, Singapore}

\author{Xiaolong Feng}
\affiliation{Research Laboratory for Quantum Materials, Singapore University of Technology and Design, Singapore 487372, Singapore}

\author{Shan Guan}
\affiliation{State Key Laboratory of Superlattices and Microstructures, Institute of Semiconductors,
Chinese Academy of Sciences, Beijing 100083, China}

\author{Wanxiang Feng}
\affiliation{Key Lab of Advanced Optoelectronic Quantum Architecture and Measurement (MOE), School of Physics,
	Beijing Institute of Technology, Beijing 100081, China}

\author{Yugui Yao}
\affiliation{Key Lab of Advanced Optoelectronic Quantum Architecture and Measurement (MOE), School of Physics,
	Beijing Institute of Technology, Beijing 100081, China}

\author{Shengyuan A. Yang}
\affiliation{Research Laboratory for Quantum Materials, Singapore University of Technology and Design, Singapore 487372, Singapore}
\affiliation{Center for Quantum Transport and Thermal Energy Science, School of Physics and Technology, Nanjing Normal University, Nanjing 210023, China}

\begin{abstract}
In a recent work, new two-dimensional materials, the monolayer MoSi$_{2}$N$_{4}$ and WSi$_{2}$N$_{4}$, have been successfully synthesized in experiment, and several other monolayer materials with the similar structure, such as MoSi$_{2}$As$_{4}$, have been predicted [{\color{blue}Science \textbf{369}, 670-674 (2020)}]. Here, based on first-principles calculations and theoretical analysis, we investigate the electronic and optical properties of monolayer MoSi$_{2}$N$_{4}$, WSi$_{2}$N$_{4}$ and MoSi$_{2}$As$_{4}$. We show that these materials are semiconductors, with a pair of Dirac-type valleys located at the corners of the hexagonal Brillouin zone. Due to the broken inversion symmetry and the effect of spin-orbit coupling, the valley fermions manifest spin-valley coupling, valley-contrasting Berry curvature, and valley-selective optical circular dichroism. We also construct the low-energy effective model for the valleys, calculate the spin Hall conductivity and the permittivity, and investigate the strain effect on the band structure. Our result reveals interesting valley physics in monolayer MoSi$_{2}$N$_{4}$, WSi$_{2}$N$_{4}$ and MoSi$_{2}$As$_{4}$, suggesting their great potential for valleytronics and spintronics applications.
\end{abstract}
	
\maketitle
\section{Introduction}
Since the discovery of graphene~\cite{novoselov2004electric,novoselov2005two,zhang2005experimental}, two-dimensional (2D) materials have been receiving tremendous attention, because of their extraordinary properties and promising applications~\cite{novoselov2005two2,bhimanapati2015recent,das2015beyond,tan2017recent}.
With the rapid advance of experimental technique, many new 2D materials beyond graphene have been realized, such as silicene and germanene~\cite{cahangirov2009,liu2011}, 2D hexagonal boron nitride~\cite{lin2010soluble,weng2016functionalized,li2016atomically}, 2D transition metal dichalcogenides~\cite{mak2010atomically,chhowalla2013chemistry,huang2013metal,tan2015two,lv2015transition}, black phosphorene~\cite{li2014black,liu2015semiconducting,eswaraiah2016black}, MXenes~\cite{naguib2011two,naguib2012two,naguib201425th}, and so on.

The progress in 2D materials also generates impetus for many other fields, such as the field of valleytronics~\cite{schaibley2016valleytronics,vitale2018valleytronics}. As an emerging degree of freedom, valley refers to the presence of multiple energy extremal points in the Brillouin zone (BZ) for low-energy carriers in a semiconductor.
Analogous to charge and spin, the valley degree of freedom can be exploited for information encoding and processing, leading to the concept of valleytronics~\cite{gunawan2006valley,rycerz2007valley,xiao2007valley,yao2008valley,gunlycke2011graphene,xiao2012coupled,zhu2012field,cai2013magnetic,jiang2013generation,xu2014spin,pan2015perfect,pan2015valley,sui2015gate,cheng2018manipulation,yu2020valley}. Although the possibility to control valley polarization had been demonstrated in certain 3D materials~\cite{zhu2012field}, it was with the advent of 2D materials that the field truly flourished. In 2D materials such as graphene and MoS$_2$-family monolayers, there are a single pair of Dirac-type valleys, forming a binary degree of freedom. Particularly, Xiao \emph{et al.}~\cite{xiao2007valley} and Yao \emph{et al.}~\cite{yao2008valley} proposed that such valleys possess contrasting geometric properties
like Berry curvature and orbital magnetic moment, which generates interesting effects such as the valley Hall effect and the optical circular dichroism. The orbital magnetic moment also allows a control of valley polarization via applied magnetic field~\cite{cai2013magnetic}. These proposals have been successfully demonstrated in experiments~\cite{mak2012control,zeng2012valley,cao2012valley,li2014valley,ju2015topological,aivazian2015magnetic,srivastava2015valley,macneill2015breaking,jiang2017zeeman,li2018valley}.

Very recently, new 2D materials of monolayer MoSi$_{2}$N$_{4}$ and WSi$_{2}$N$_{4}$ were synthesized in experiment~\cite{hong2020chemical}. It was found that they are semiconductors with high strength and excellent ambient stability.  Based on calculations, a family of 2D materials with the similar structure, such as monolayer MoSi$_{2}$As$_{4}$, have also been predicted~\cite{hong2020chemical,wang2020structure}.
Motivated by the discovery of these new materials, in this work, based on first-principles calculations and theoretical modeling, we investigate the electronic and optical properties of monolayer MoSi$_{2}$N$_{4}$, WSi$_{2}$N$_{4}$ and MoSi$_{2}$As$_{4}$, focusing on the valleytronic aspects. We show that in terms of valley structures, these materials are very similar to the MoS$_2$-family monolayers. They have a pair of Dirac-type valleys located at the corners of the hexagonal BZ, connected by the time reversal symmetry. With the broken inversion symmetry and sizable spin-orbit coupling (SOC), the two valleys exhibit contrasting features in spin splitting, Berry curvature, and optical circular dichroism. This means the valley polarization in these materials can also be effectively controlled by transport, optical, and magnetic means. In addition, we show that the lattice strain can be utilized to tune the band structures, and may drive transitions between direct and indirect band gaps in these materials.
Given that these 2D materials have gap sizes complementary to the 2D MoS$_2$-family materials, our results here suggest their great potential for valleytronic and spintronic applications.

\begin{figure}[t]
	\includegraphics[width=8.6cm]{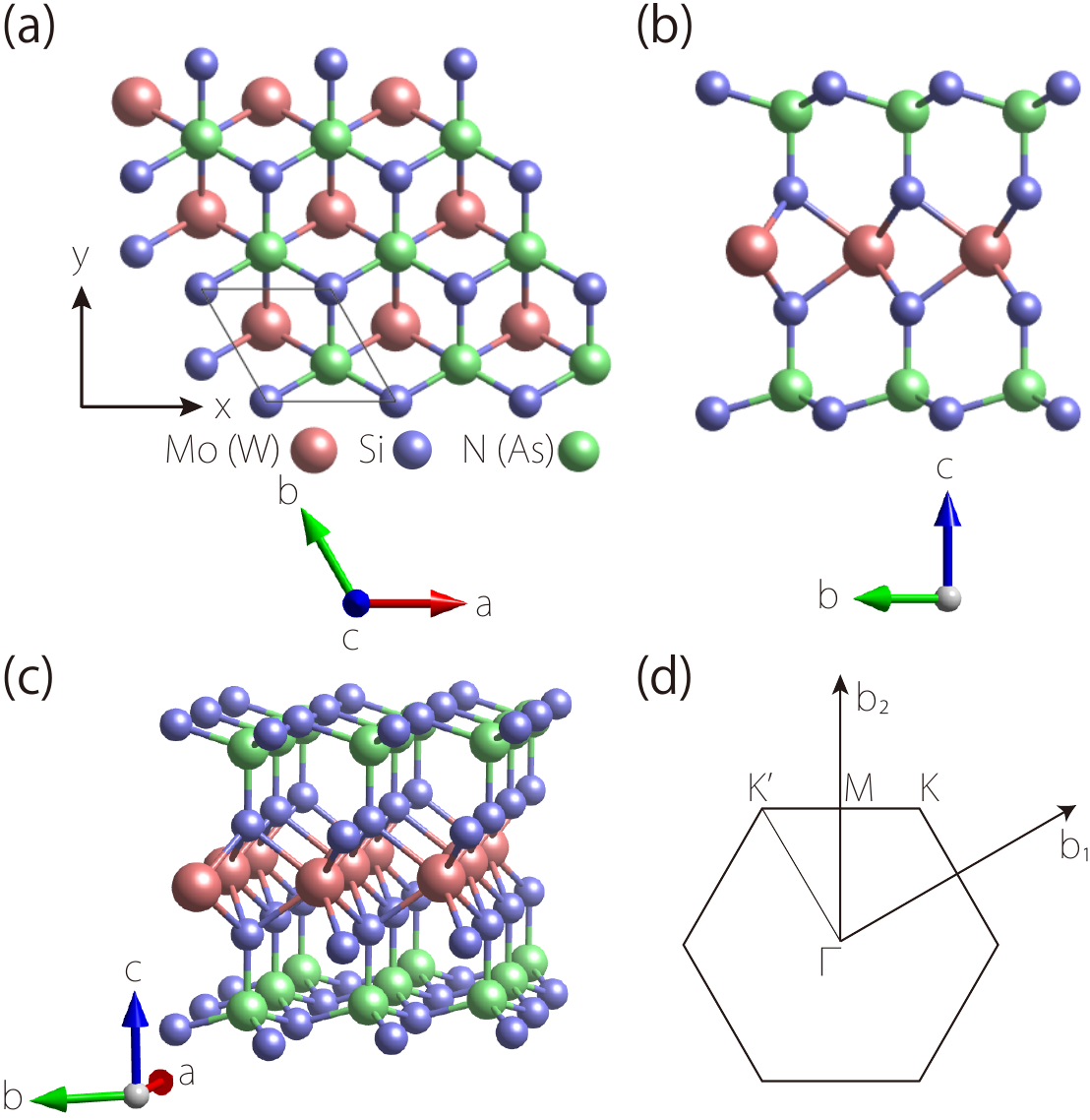}
	\caption{ (a) Top view, (b) side view, and (c) perspective view of the crystal structure of the studied monolayer material. The primitive cell is shown with the solid line in (a). (d) Brillouin zone with the high-symmetry points labeled. }
	\label{fig1}
\end{figure}

\section{Computation METHODS}

Our first-principles calculations were based on the density functional theory (DFT), using the projector augmented wave method as implemented in the Vienna \emph{ab initio} simulation package~\cite{Kresse1994,Kresse1996,PAW}. The generalized gradient
approximation with the Perdew-Burke-Ernzerhof (PBE)~\cite{PBE} realization was adopted for the exchange-correlation functional. The band structure result was further checked by using the more accurate Heyd-Scuseria-Ernzerhof hybrid functional method (HSE06)~\cite{heyd2003hybrid} (see Appendix~\ref{A}). The cutoff energy was set as 500 eV. The energy and force convergence criteria were set to be $10^{-6}$ eV and $0.001$ eV/\AA, respectively. The BZ was sampled with a $\Gamma$-centered $k$ mesh of size $15\times 15\times 1$. A vacuum layer with a thickness of 20 \AA\ was taken to avoid artificial interactions between periodic images. The PyProcar code~\cite{herath2020pyprocar} was used to plot the band structure with the spin polarization. We construct Wannier functions by projecting the Bloch states onto atomic-like trial orbitals without an iterative procedure~\cite{Marzari1997,Souza2001,mostofi2008wannier90}. In the calculation of the intrinsic spin Hall conductivity [Eq.~(\ref{shc})], a dense $k$-point mesh of size $1400 \times 1400 \times 1$ was adopted.
And for the permittivity calculation, a $k$-point mesh with size $36\times 36\times 1$ was used.
The permittivities are suitably renormalized to exclude the vacuum region by using the experimental value 10.7 \AA~as the effective thickness~\cite{hong2020chemical}.

\section{electronic band and valley structures}\label{EBS}

\begin{figure*}[htbp]
	\includegraphics[width=17.8cm]{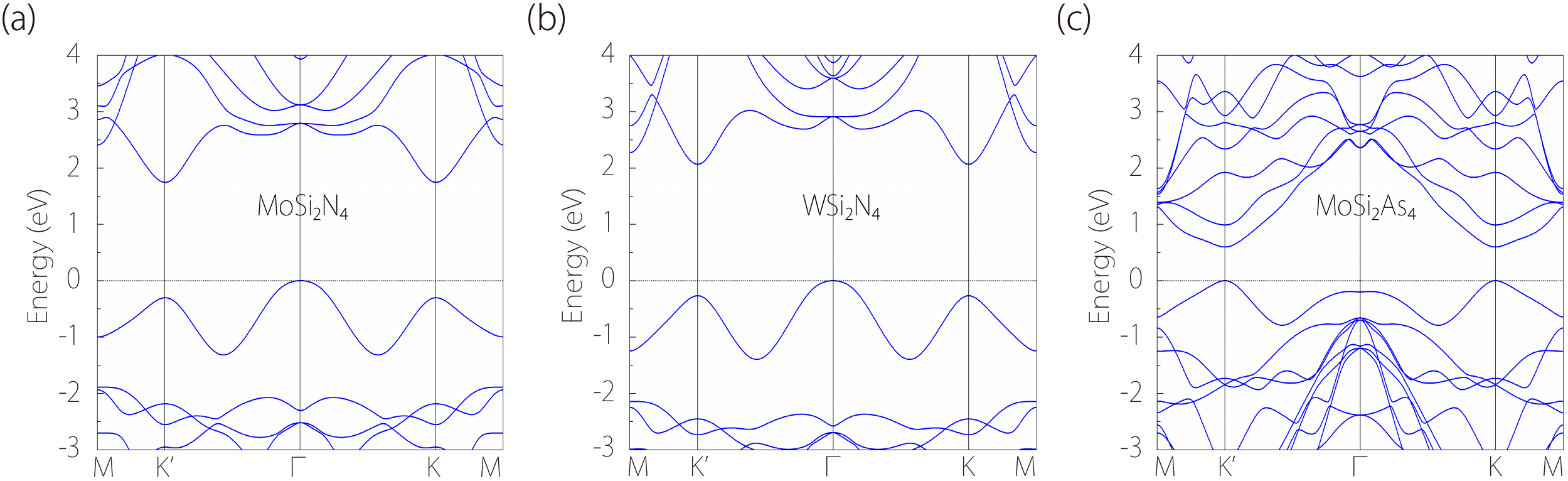}
	\caption{ Band structure of monolayer (a) MoSi$_{2}$N$_{4}$, (b) WSi$_{2}$N$_{4}$, and (c) MoSi$_{2}$As$_{4}$ in the absence of SOC. }
	\label{fig2}
\end{figure*}

\begin{figure}[htbp]
	\includegraphics[width=7.6cm]{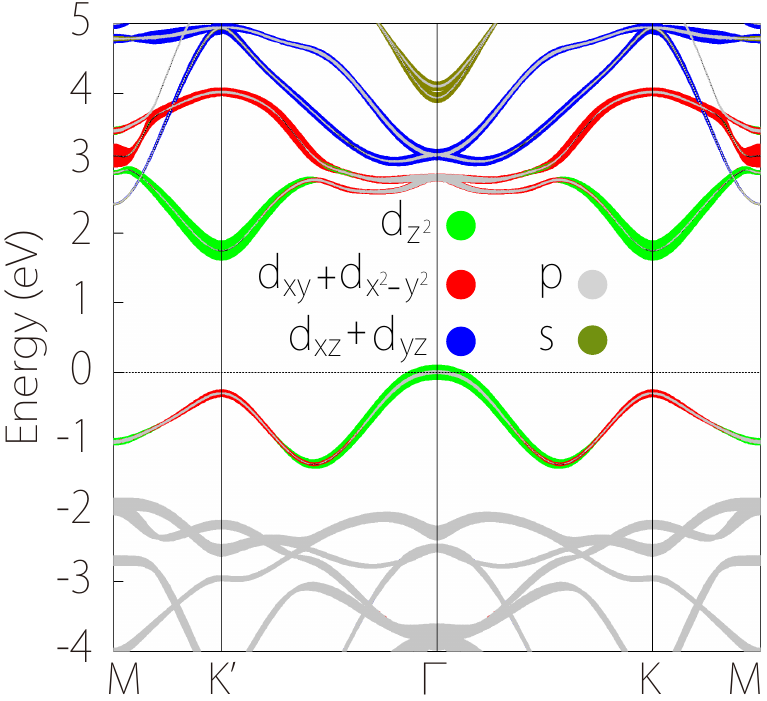}
	\caption{Orbital projected band structure for monolayer MoSi$_{2}$N$_{4}$. The symbol size is proportional to the weight projected to a particular orbital. }
	\label{fig3}
\end{figure}

\begin{figure*}[htbp]
	\includegraphics[width=17.8cm]{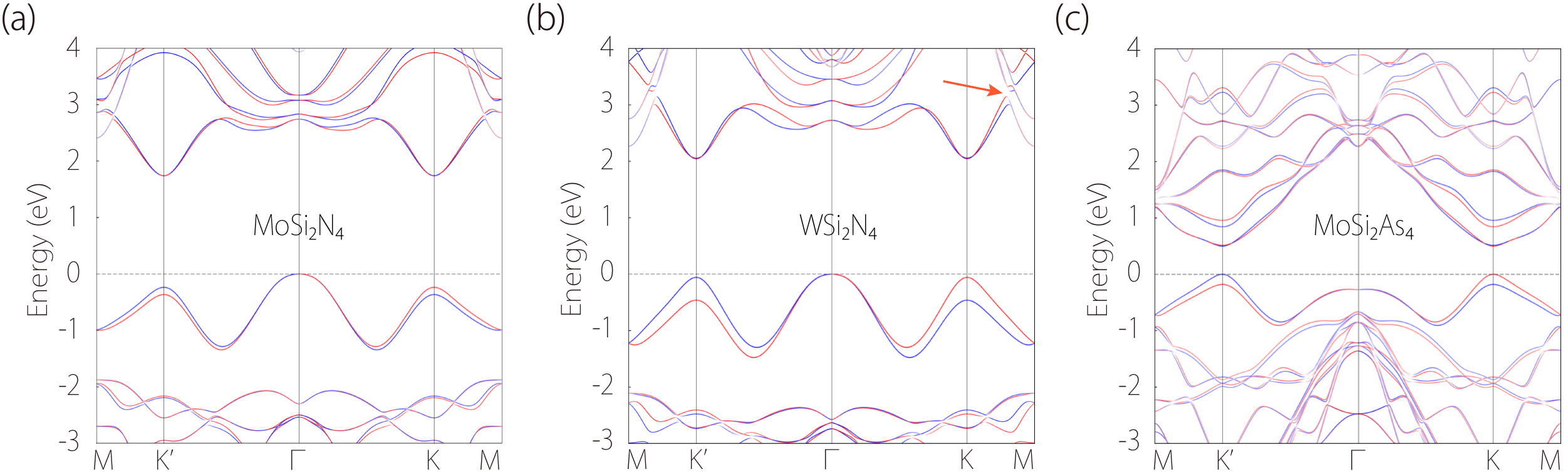}
	\caption{Band structures of monolayer (a) MoSi$_{2}$N$_{4}$, (b) WSi$_{2}$N$_{4}$, and (c) MoSi$_{2}$As$_{4}$ in the presence of SOC. The red and blue colors indicate the spin-up and spin-down bands, respectively. The red arrow in (b) indicates the band near-degeneracy, which causes the sharp features in Fig.~6(b).}
	\label{fig4}
\end{figure*}

The monolayer MoSi$_{2}$N$_{4}$ and WSi$_{2}$N$_{4}$ have been experimentally realized by the chemical vapor deposition (CVD) method~\cite{hong2020chemical}. These two materials and MoSi$_{2}$As$_{4}$ share the same type of  hexagonal lattice structure with space group $P \overline{6} m 2$ (No.~$187$). As shown in Fig.~\ref{fig1}(a)-\ref{fig1}(c), the structure is built up
by septuple atomic layers in the sequence of N(As)-Si-N(As)-Mo(W)-N(As)-Si-N(As). It is important to note that the structure breaks the inversion symmetry, but preserves a horizontal mirror $M_z$ corresponding to the plane of the Mo (W) layer.  %%
From our first-principles calculation, the fully
optimized lattice parameters for monolayer MoSi$_{2}$N$_{4}$, WSi$_{2}$N$_{4}$, and MoSi$_{2}$As$_{4}$ are $a =b= 2.909$ \AA,  $2.912$ \AA, and  $3.621$ \AA, respectively. These parameters agree well with the previously reported values~\cite{hong2020chemical}.

We then investigate the electronic band structures of these three materials. We first calculate the band structures in the absence of SOC. The obtained results are plotted in Fig.~\ref{fig2}. One observes that MoSi$_{2}$N$_{4}$ and WSi$_{2}$N$_{4}$ are indirect band-gap semiconductors with the conduction-band minimum (CBM) at $K$/$K'$ and valence-band maximum (VBM) at  $\Gamma$. Here, $K$ and $K'$ are the two inequivalent corners of the hexagonal BZ [see Fig.~\ref{fig1}(d)]. In contrast, MoSi$_{2}$As$_{4}$ is a direct band-gap semiconductor with both CBM and VBM at $K$/$K'$.
To understand the composition of the low-energy states, we project the states to atomic orbitals of the constituent atoms. The result for
MoSi$_{2}$N$_{4}$ is shown in Fig.~\ref{fig3}. The important observation is that the states near the band edges are dominated by the Mo-$d$ orbitals. More specifically, the states around CBM at $K$/$K'$ are dominated by the Mo $d_{z^2}$ orbital. As for the highest valence band, the states around $\Gamma$ are also dominated by the Mo $d_{z^2}$ orbital, but the states around $K$/$K'$ are dominated by
the Mo $d_{xy}$ and $d_{x^2-y^2}$ orbitals.  This orbital character is also shared by the other two materials WSi$_{2}$N$_{4}$ and MoSi$_{2}$As$_{4}$.

Next, we consider the band structures with SOC included, and the results are plotted in Fig.~\ref{fig4}. One observes that MoSi$_{2}$N$_{4}$ and WSi$_{2}$N$_{4}$ are still indirect band-gap semiconductors, and MoSi$_{2}$As$_{4}$ still retains a direct band gap. The obtained band gap values with PBE method are listed in Table~\ref{table1}. We have also use the more sophisticated hybrid functional method (HSE06)~\cite{heyd2003hybrid} to check the band structure results. We find that the main band features are consistent between these two methods, except that for WSi$_{2}$N$_{4}$, HSE06 predicts a direct band gap with the valence state at $K$/$K'$ about 5 meV higher than that of the $\Gamma$ point. The band gap values from HSE06 approach are also listed in Table~\ref{table1} for comparison. The corrections of HSE06 on PBE band gaps are 0.556 eV, 0.564 eV, and 0.191 eV for monolayer MoSi$_2$N$_4$, WSi$_2$N$_4$, and MoSi$_2$As$_4$, respectively. These values are lower than the corresponding corrections for some bulk semiconductors like GaN, Si, and GaAs~\cite{heyd2005energy}, but they are comparable to those for the 2D MoS$_2$-family materials~\cite{ramasubramaniam2012large}.

Compared with Fig.~\ref{fig2}, one observes that the main effect of SOC in Fig.~\ref{fig4} is to cause a spin splitting of the original bands. Here, we use the red/blue color to indicate the spin polarization $\left\langle n\bm{k}\left|\hat{s}_{z}\right| n\bm{k}\right\rangle$ of the band eigenstate $|n\bm{k}\rangle$. From the symmetry perspective, the spin splitting is a direct consequence of the inversion symmetry breaking, which lifts the spin degeneracy at each generic $k$ point. Moreover, due to the existence of the horizontal mirror $M_z$ which operates on spin as $-i\hat{s}_z$, the band eigenstates must also be eigenstates of the $\hat{s}_z$ operator, i.e., they are fully spin-polarized in the out-of-plane direction. Lastly, the time reversal symmetry $T$ is preserved in the system, so that the spin polarizations at $K$ and $K'$ are opposite, as the two points are connected by the $T$ operation. This is particularly evident when looking at the highest valence band in Fig.~\ref{fig4}.

From the band structures in Fig.~\ref{fig4}, another important observation is that there exists a valley degree of freedom. All three materials have conduction band valleys at $K$ and $K'$. For monolayer MoSi$_{2}$As$_{4}$, its valence band also has valleys at $K$ and $K'$, whereas for the other two materials, although the VBM occurs at $\Gamma$, the $K$/$K'$ valleys are still well defined and not far in energy. As we have mentioned, $K$ and $K'$ points are connected by the time reversal operation, so any property that is odd under $T$ will have opposite values between the $K$ and $K'$ valleys. The spin polarization we just analyzed is one such example. In the following, we shall examine other properties, and these will be useful for controlling the valley degree of freedom, as has been originally proposed in Refs.~\cite{xiao2007valley,yao2008valley,cai2013magnetic}.  Note that the exact gap values will not affect the main features of the valley properties, so in the following, we shall base our discussion mostly on the PBE band structures.

 \begin{table}
	\caption{Summary of key band features obtained from DFT calculations (SOC included). $E_g$ is global band gap, $E_K$ is the energy gap at valley $K$/$K'$, and $\Delta_s$ is the spin splitting of the valence band at $K$/$K'$. The unit is eV.}
	\begin{ruledtabular}
		\begin{tabular}{lcccc}	
		\qquad\qquad\qquad &	Method & $E_g$ & $E_K$ & $\Delta_s$ \\
			
		\hline\hline MoSi$_{2}$N$_{4}$&  PBE& 1.742&1.976& 0.129  \\
	 	   \qquad\qquad\qquad&	 HSE06& 2.298& 2.442& 0.172  \\		               	
			
	        \hline 	WSi$_{2}$N$_{4}$&   PBE & 2.044& 2.087& 0.404  \\
		\qquad\qquad\qquad&   HSE06 & 2.608& 2.608& 0.513  \\
			
		    \hline 	MoSi$_{2}$As$_{4}$&  PBE & 0.500& 0.500& 0.180  \\
		\qquad\qquad\qquad&    HSE06& 0.691& 0.691& 0.284  \\
		\end{tabular}
	\end{ruledtabular}
	\label{table1}
\end{table}

\section{effective model}

To better understand the valley properties, we shall first construct an effective model which describes the valley structures at $K$ and $K'$ points.
As analyzed in Sec.~\ref{EBS}, the conduction band at $K$/$K'$ is dominated by the $d_{z^2}$ orbital, whereas the valence band at $K$/$K'$ is dominated by the  $d_{xy}$ and $d_{x^2-y^2}$ orbitals. This feature is similar to the case in monolayer MoS$_2$. At $K$ and $K'$, the little group is $C_{3h}$ which may be generated by the symmetry elements $C_{3}$ and $M_z$. By choosing the two orbital basis $\left|\psi_{c}^{\tau}\right\rangle=\left|d_{z^{2}}\right\rangle$ and $\left|\psi_{v}^{\tau}\right\rangle=\frac{1}{\sqrt{2}}\left(\left|d_{x^{2}-y^{2}}\right\rangle+i \tau\left|d_{x y}\right\rangle\right)$, where $\tau=\pm 1$ is the valley index corresponding to $K$/$K'$, the $k\cdot p$ effective model at $K$/$K'$ in the absence of SOC can be obtained as (expanded to linear order in $k$)
\begin{equation}\label{kp1}
{H}_{0}^\tau=\alpha\left(\tau k_{x} {\sigma}_{x}+k_{y} {\sigma}_{y}\right)+\frac{\Delta}{2} {\sigma}_{z},
\end{equation}
where $k$ is measured from $K$/$K'$, $\sigma$'s are the Pauli matrices denoting the two-orbital degree of freedom, $\alpha$ and $\Delta$ are real valued model parameters.

Then, SOC is taken into account as a perturbation to the Hamiltonian in Eq.~(\ref{kp1}). Approximating the SOC term as an intra-atomic contribution and keeping only the leading order (in $k$) term, we obtain the effective model in the presence of SOC as
\begin{equation}\label{kp2}
{H}^\tau=\alpha\left(\tau k_{x} {\sigma}_{x}+k_{y} {\sigma}_{y}\right)+\frac{\Delta}{2} {\sigma}_{z}-\frac{\lambda}{2} \tau (\sigma_{z}-1) {s}_{z},
\end{equation}
where $s_z$ is the Pauli matrix for spin and $\lambda$ represents the effective SOC strength. Compared with the band structures in Fig.~\ref{fig4}, one observes that the SOC term in (\ref{kp2}) captures the key feature of spin splitting at $K$/$K'$: the splitting is pronounced in the valence band (with a value of $2\lambda$), whereas it is very small for the conduction band. In addition, the SOC term shows that the spins are fully polarization in the out-of-plane direction due to the $M_z$ symmetry, as we mentioned before. We have used this model to fit the DFT band structures. The extracted model parameters are listed in Table~\ref{table2}.

The effective model (\ref{kp2}) is diagonal in the real spin. For each spin channel $s_z=\pm 1$, the model takes a form of a gapped Dirac model. This demonstrates that the valleys at $K$ and $K'$ here are Dirac-type valleys~\cite{yang2016dirac}.
One can observe that model (\ref{kp2}) has essentially the same form as that for the monolayer MoS$_2$-family materials~\cite{xiao2012coupled}. This is because both classes of materials have similar low-energy band structures at $K$/$K'$, and they possess the same symmetry group. Thus, regarding the valley structure, these three materials are very similar to the monolayer MoS$_2$-family materials. Nevertheless, there are also differences between them. For example, the band gap in MoSi$_{2}$As$_{4}$ is smaller than the MoS$_2$-family materials (typically 1.5 to 2 eV), and the gaps in MoSi$_{2}$N$_{4}$ and WSi$_{2}$N$_{4}$ are larger, so they offer an energy coverage complementary to the MoS$_2$-family materials.

\begin{table}
	\caption{ Effective model parameters extracted from the first-principles band structures. The unit is eV$\cdot$\AA\ for $\alpha$, and eV for $\Delta$ and $\lambda$. }
	\begin{ruledtabular}
		\begin{tabular}{lcccc}	
	\qquad\qquad\qquad &	Method & $\alpha$  & $\Delta$ & $\lambda$ \\
	
	\hline\hline MoSi$_{2}$N$_{4}$&  PBE& 4.021&2.041& 0.065  \\
            	\qquad\qquad\qquad&	 HSE06& 5.080& 2.528& 0.086  \\		               	
	
	\hline 	WSi$_{2}$N$_{4}$&   PBE &  4.756& 2.289& 0.202  \\
     	\qquad\qquad\qquad&   HSE06 & 5.876& 2.865& 0.257  \\
	
	\hline 	MoSi$_{2}$As$_{4}$&  PBE & 2.244& 0.590& 0.090  \\
	    \qquad\qquad\qquad&    HSE06& 3.102& 0.833& 0.142  \\			
		\end{tabular}
	\end{ruledtabular}
	\label{table2}
\end{table}

\section{Valley-contrasting Berry curvature and circular dichroism}
Dirac-type valleys possess interesting physical properties such as Berry curvature and optical circular dichroism. It is important to note that these properties are odd under $T$, so they will take opposite values for the two valleys $K$ and $K'$.

Let's first consider the Berry curvature. For 2D systems, it only has a $z$ component, which behaves as a pseudoscalar. For a state $|n\bm k\rangle$, its Berry curvature can be calculated by
\begin{equation}
\Omega_{n\bm k}=-2 \operatorname{Im} \sum_{n'\neq n} \frac{\left\langle n \bm{k}\left|v_{x}\right| n' \bm{k}\right\rangle\left\langle n' \bm{k}\left|v_{y}\right| n \bm{k}\right\rangle}{(\omega_{n^{\prime}}-\omega_{n})^{2}},
\end{equation}
where the $v$'s are the velocity operators, and $E_n=\hbar\omega_{n}$ is the energy of the state $|n\bm k\rangle$. By using the effective model (\ref{kp2}), one finds that the Berry curvature for the two conduction bands (with opposite $s_z$) at $K$/$K'$ valley is given by
\begin{equation}\label{BC}
\Omega_{c}^\tau(\boldsymbol{k},s_z)=-\tau \frac{2 \alpha^{2} \tilde{\Delta}_{\tau s_z}}{\left(\tilde{\Delta}_{\tau s_z}^{2}+4 \alpha^{2} k^{2}\right)^{3/2}},
\end{equation}
where $\tilde{\Delta}_{\tau s_z} = \Delta-\tau s_{z} \lambda$. And for the two spin-split valence bands at $K$/$K'$, we have
\begin{equation}
  \Omega_v^\tau (\bm k, s_z)=-\Omega_c^\tau (\bm k, s_z).
\end{equation}
It is clearly from Eq.~(\ref{BC}) that the two valleys have nonzero and opposite Berry curvatures. The magnitude of the Berry curvature is peaked at the valley center $k=0$.

\begin{figure*}[t]
	\includegraphics[width=17.8cm]{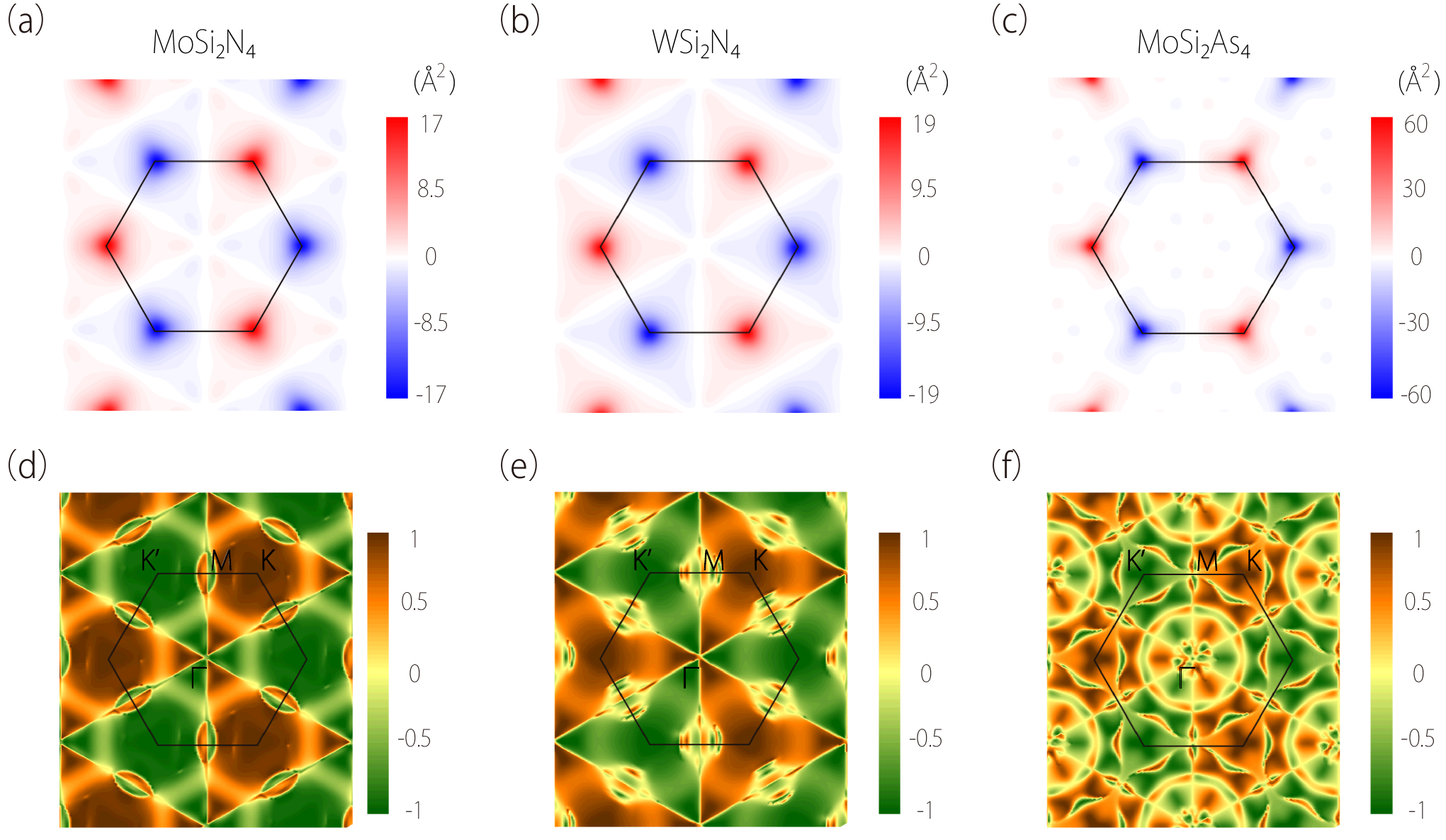}
	\caption{ Distribution of Berry curvature summed for all valance bands for monolayer (a) MoSi$_{2}$N$_{4}$, (b) WSi$_{2}$N$_{4}$, and (c) MoSi$_{2}$As$_{4}$. (d)-(e) show the corresponding circular polarization $\eta$ calculated for transitions between the highest valence band and lowest conduction band. The black lines indicate the BZ.}
	\label{fig5}
\end{figure*}

\begin{figure}[t]
	\includegraphics[width=8.6cm]{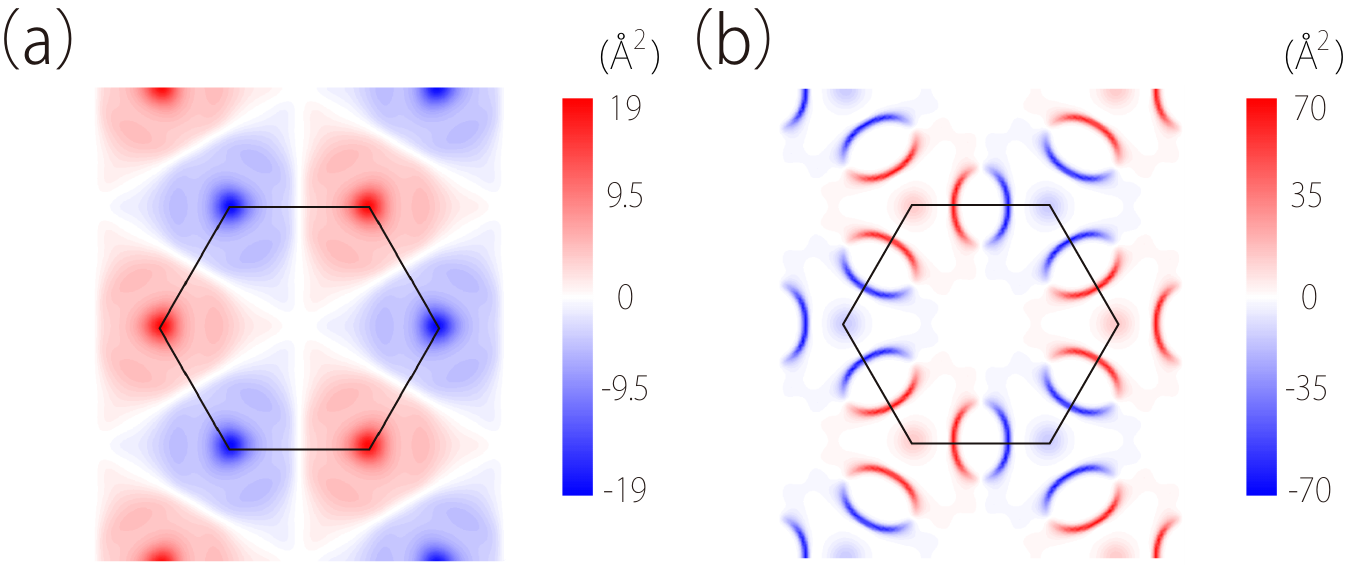}
	\caption{Distribution of the combined Berry curvature for (a) the two highest valence bands, and (b) the two lowest conduction bands for monolayer WSi$_{2}$N$_{4}$. The black lines indicate the BZ.}
	\label{fig6}
\end{figure}

In Fig.~\ref{fig5} and Fig.~\ref{fig6}, we have shown the Berry curvature distribution obtained from DFT calculations, which go beyond the effective model. In Fig.~\ref{fig5}, we plot the total Berry curvature for all the valence bands $\Omega(\bm k)=\sum_{n\in occ.} \Omega_{n\bm k}$ for the three materials. One clearly observes that the distribution is peaked and takes opposite values at $K$ and $K'$ valleys. In Fig.~\ref{fig6}(b) [Fig.~\ref{fig6}(a)], we also plot the combined Berry curvature for the two conduction (valence) bands closest to the band gap for monolayer WSi$_{2}$N$_{4}$ (results for the other two are similar). The valley-contrasting character is also faithfully reproduced. We note that in Ref.~\cite{wang2020structure}, Berry curvature distribution with similar
valley features was also found in monolayer $\alpha_{2}-\mathrm{WSi}_{2} \mathrm{P}_{4}$. In Fig.~\ref{fig6}(b), one observes there are additional sharp features between $K$/$K'$ and $M$. These are at higher energy and due to the band near-degeneracies as marked in Fig.~\ref{fig4}(b).

\begin{figure}[t]
	\includegraphics[width=8.6cm]{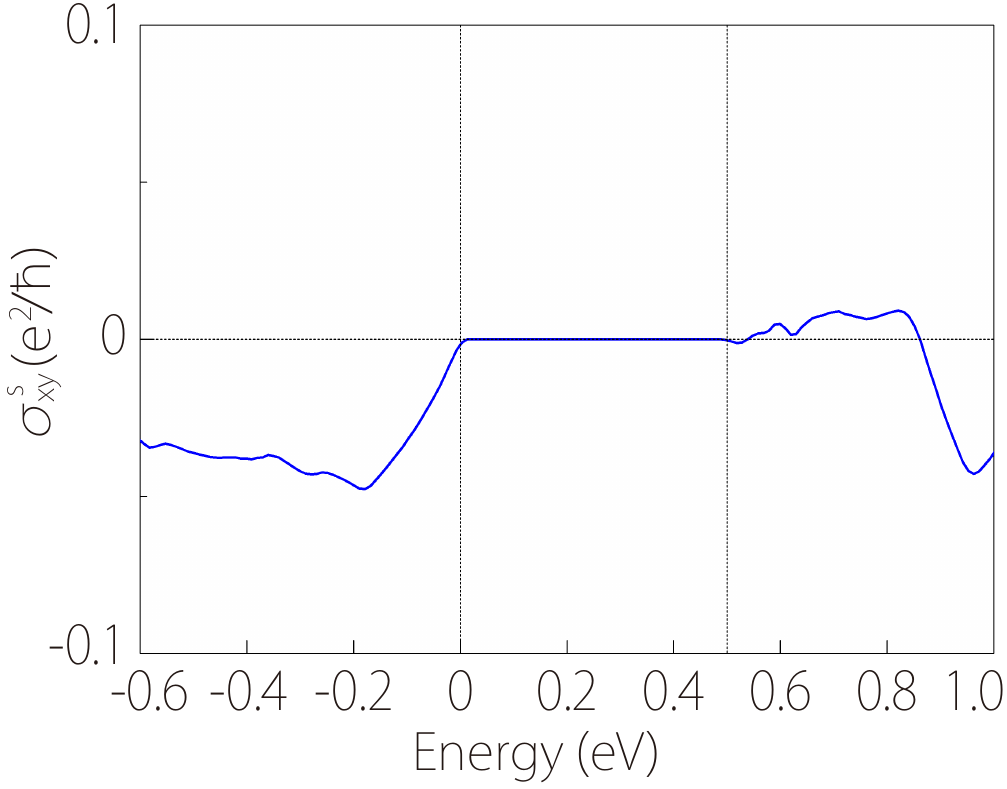}
	\caption{Calculated intrinsic spin Hall conductivity $\sigma_{x y}^{s}$ for monolayer MoSi$_{2}$As$_{4}$. Dashed vertical lines indicate the band edge locations.}
	\label{fig7}
\end{figure}

Such nonzero Berry curvature is the key ingredient for the anomalous Hall transport, which means a longitudinal (in-plane) $E$ field will deflect carriers in transverse directions without applied magnetic field. This is most readily understood in the semiclassical formalism~\cite{xiao2010berry}, where the Berry curvature directly enters into the equations of motion and generates a so-called anomalous velocity term $\propto \bm E\times\bm{\Omega}$. For our current systems, there is no net charge current at linear order because the $T$ symmetry is preserved. However, there could exist valley and spin Hall effect. The valley Hall effect occurs in these materials immediately after electron doping. As for hole doping, in MoSi$_{2}$N$_{4}$ and WSi$_{2}$N$_{4}$, one has to reach a doping level with $K$/$K'$ valleys occupied to observe the effect.
The case with hole doping in MoSi$_{2}$As$_{4}$ could be more interesting. There, the VBM occurs at $K$ and $K'$, and the valleys are spin polarized. The valley Hall effect is therefore associated with a pronounced spin Hall effect. We have evaluated the intrinsic spin Hall conductivity for the doped monolayer MoSi$_{2}$As$_{4}$, defined by
\begin{equation}\label{shc}
  \sigma_{xy}^s=\sum_n\frac{e^2}{\hbar}\int\frac{d^2k}{(2\pi)^2}f_{n\bm k}(s_z)_{n\bm k} \Omega_{n\bm k},
\end{equation}
where $f_{n\bm k}$ is the equilibrium distribution function for the doped carriers, $(s_z)_{n\bm k}$ is the $s_z$ spin polarization of the state $|n\bm k\rangle$, and we have included an extra factor of $2e/\hbar$ in the definition such that the magnitude of $\sigma_{xy}^s$ can be better compared with charge conductivities. The result obtained from the DFT calculation is plotted in Fig.~\ref{fig7}. One observes that $\sigma_{xy}^s$ takes a negative value for hole doping and switches sign for electron doping (and again there is a sign change around 0.8 eV). The magnitude of the spin Hall conductivity ($\sim 0.05 e^2/\hbar$) under hole doping is comparable to that of MoS$_2$-family materials~\cite{feng2012intrinsic}. We have estimated that to achieve the peak magnitude in Fig.~\ref{fig7} (at about $-0.18$ eV), the hole doping concentration is around $1 \times 10^{14} \mathrm{~cm}^{-2}$, which can be achieved by the current experimental technique~\cite{efetov2010controlling, ye2011accessing}. The Hall effect will results in both valley and spin polarizations at the edges along the transverse direction, which can be respectively detected by the optical circular dichroism and the magneto-optical Kerr effect.

The nonzero Berry curvature indicates the Dirac-type valleys have a sense of chirality, which can manifest in the effect of optical circular dichroism~\cite{yao2008valley}. It has been shown that for a general two-band system, the degree of circular polarization $\eta$ is proportional to the Berry curvature~\cite{liu2018circular}. For our current systems, at each valley there are four low energy bands. As optical transitions are vertical and preserve the spin, they can only happen between a pair of bands belonging to the same spin channel. Hence, one can consider the two spin channels separately.

The circular polarization for optical transition between a pair of states is defined by
\begin{equation}
\eta=\frac{\left|\mathcal{M}_{+}\right|^{2}-\left|\mathcal{M}_{-}\right|^{2}}{\left|\mathcal{M}_{+}\right|^{2}+\left|\mathcal{M}_{-}\right|^{2}},
\end{equation}
where $\mathcal{M}_{\pm}$ is the optical transition matrix element for incident light with $\sigma\pm$ circular polarizations. As
$\mathcal{M}_{\pm}=\frac{1}{\sqrt{2}}(\mathcal{M}_x\pm i\mathcal{M}_y)$, and $\mathcal{M}_{x/y}=m_e\langle c\bm k|v_{x/y}|v\bm k\rangle$ ($m_e$ is the free electron mass), the circular polarization can be readily evaluated in the effective model or by DFT calculations.

Consider the transition from the topmost valence band (which has a definite spin polarization) to the conduction band at $K$/$K'$ valleys. Using the effective model in Eq.~(\ref{kp2}), we find that the corresponding $k$-resolved circular polarization is given by
\begin{equation}
\begin{aligned}
  \eta^\tau(\bm k)=\frac{\tau (\Delta-\lambda) \sqrt{(\Delta-\lambda)^2+4 \alpha^{2} k^{2}}}{(\Delta-\lambda)^2+2 \alpha^{2} k^{2}}.
\end{aligned}
\end{equation}
At the valley center $\bm k=0$, we have $\eta^\tau=\tau$, indicating that the transition at the valley center is exclusively coupled to a specific circular polarization.

The corresponding results obtained from DFT calculations for the three materials are shown in Fig.~\ref{fig5}(d-f). Here, the plotted $\eta(\bm k)$ is for the transitions between the highest valence band and the lowest conduction band, over the whole BZ. Note that these transitions are at different light frequencies. The lowest transition frequency occurs at the two valleys, where $\eta$ takes opposite signs, consistent with our analysis using the effective model.

\begin{figure*}[htbp]
	\includegraphics[width=17.8cm]{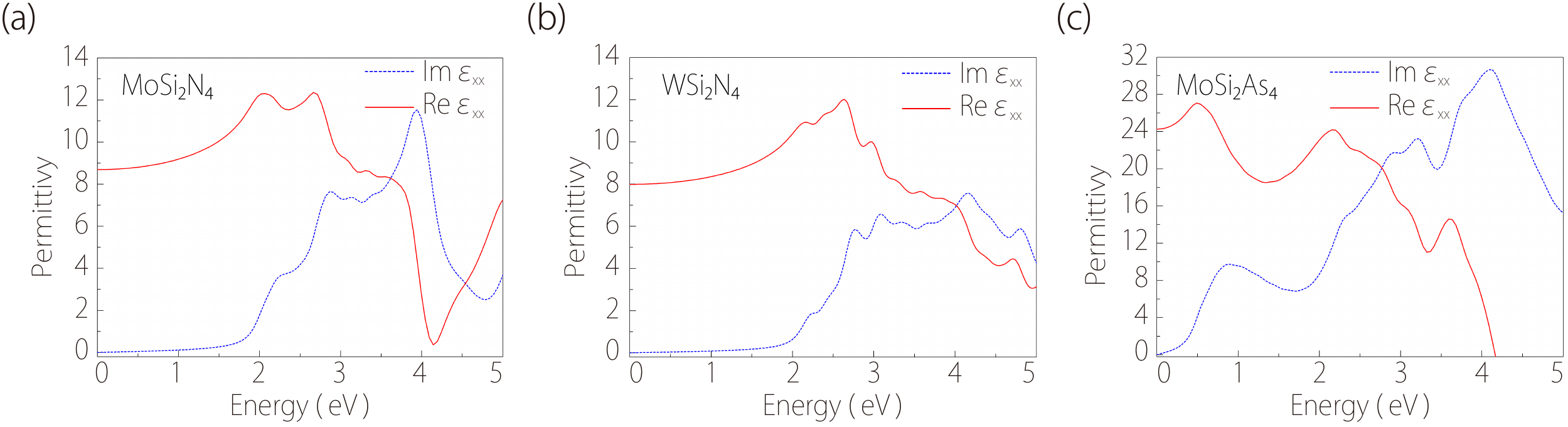}
	\caption{ Calculated real and imaginary parts of the permittivity of monolayer (a) MoSi$_{2}$N$_{4}$, (b) WSi$_{2}$N$_{4}$, and (c) MoSi$_{2}$As$_{4}$.}
	\label{fig8}
\end{figure*}

To assess the ability of optical absorption of these materials, we have also calculated their permittivity. Due to the $D_{3h}$ symmetry, the permittivity tensor must be isotropic in plane. In Fig.~\ref{fig8}, we plot the real and imaginary parts of $\varepsilon_{xx}$ calculated by DFT approach. The imaginary part $\text{Im}(\varepsilon_{xx})$ is related to the optical absorption. {The absorption edges in $\text{Im}(\varepsilon_{xx})$ reflect the semiconducting band gaps.} One observes that the absorption in MoSi$_{2}$As$_{4}$ in the visible light range is around two times stronger than in MoSi$_{2}$N$_{4}$ and WSi$_{2}$N$_{4}$. The optical absorption strength here is comparable to that of monolayer MoS$_2$~\cite{ahmed2015bonding,lahourpour2019structural,ermolaev2020broadband}. The real part $\text{Re}(\varepsilon_{xx})$ also manifests features of a semiconductor: it is positive in the low energy range, and below the band gap it directly determines the refractive index of the material.

\section{Strain effect on band structure}
\begin{figure*}[htbp]
	\includegraphics[width=17.8cm]{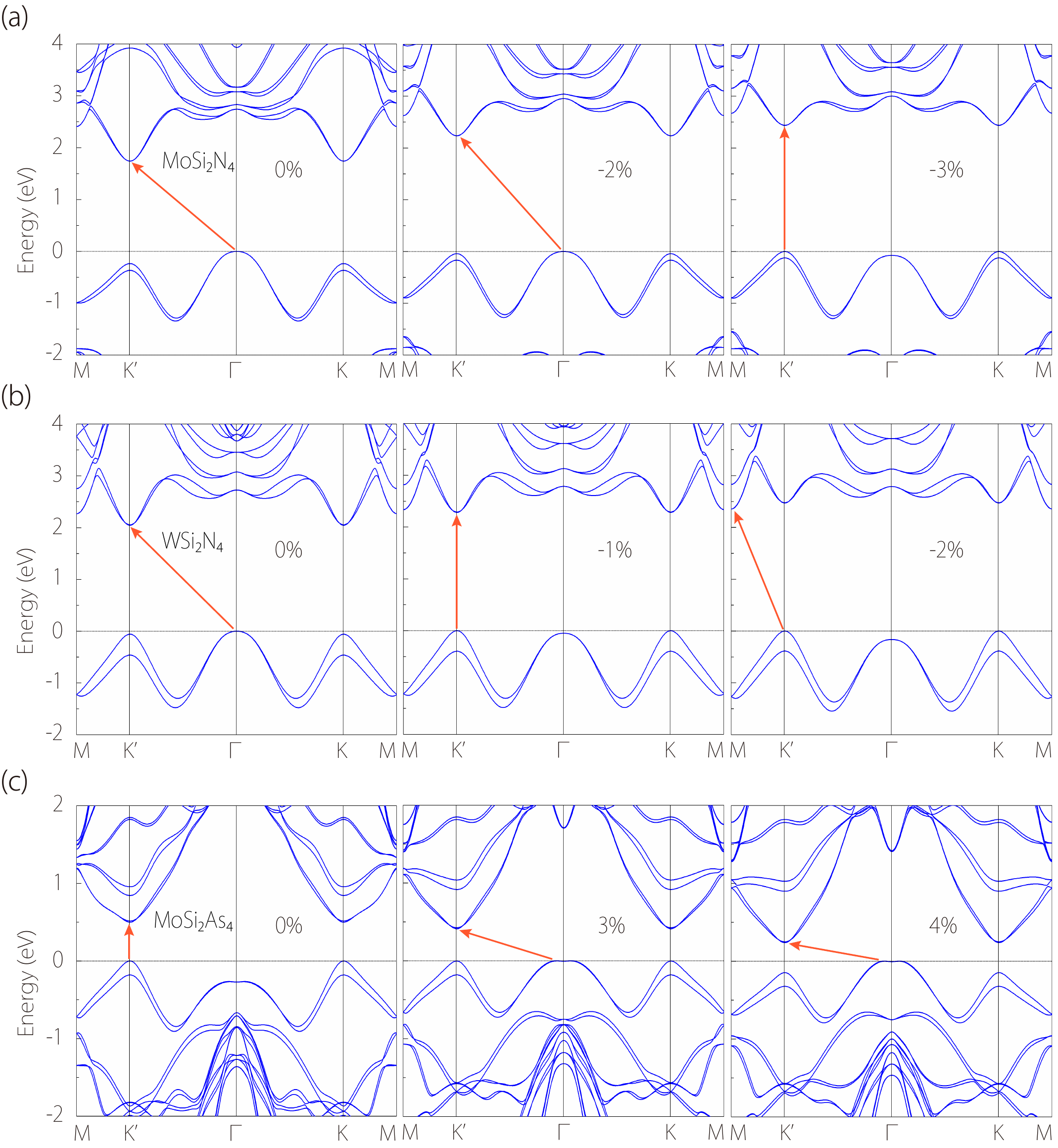}
	\caption{Strain effect on the band structures (SOC included) of monolayer (a) MoSi$_{2}$N$_{4}$, (b) WSi$_{2}$N$_{4}$, and (c) MoSi$_{2}$As$_{4}$. }
	\label{fig9}
\end{figure*}

2D materials can usually sustain large lattice strains, and their properties can be effectively tuned by strain engineering.
Here, we study the effect of biaxial strains on the band structures of the three materials. The most important finding is that a moderate strain ($<5\%$) can generate a transition between direct and indirect band gaps in these materials. As shown in Fig.~\ref{fig9}, for monolayer
MoSi$_{2}$N$_{4}$ and WSi$_{2}$N$_{4}$, this transition occurs at small compressive strain of $-3\%$ and $-1\%$, respectively. During the transition, the VBM switches from $\Gamma$ to $K$/$K'$ points. Interestingly, in  WSi$_{2}$N$_{4}$, there is a second transition around $-2\%$, during which the VBM is unchanged (at $K$/$K'$), but the CBM is switched from $K$/$K'$ to $M$. Meanwhile, the transition in monolayer MoSi$_{2}$As$_{4}$ occurs at tensile strain around $3\%$, at which the VBM changes from $K$/$K'$ to $\Gamma$. Thus, in these materials, the low-energy band features are rather sensitive to strain. A common feature is that with increasing lattice parameter, the valence band at $\Gamma$ tends to shift up whereas the valence band at $K$/$K'$ valleys tends to move downward. This leads to the transition between indirect and direct band gaps. This feature is also verified by the more sophisticated HSE06 approach: Although the exact strain values for the transitions generally differ between different computational methods, the qualitative results still remain valid. We point out that this strain-induced transition
 could be useful for valley switching, i.e., the applied strain can be used to control the emergence/disappearance of the valley degree of freedom.

\section{Discussion and Conclusion}

In this work, we have studied the emerging valley degree of freedom in the newly synthesized 2D materials MoSi$_{2}$N$_{4}$ and WSi$_{2}$N$_{4}$, and the closely related MoSi$_{2}$As$_{4}$. We have shown that their valley structure is very similar to the MoS$_2$-family materials. And we have also pointed out their noticeable differences, such as their different gap values complementary to the MoS$_2$-family materials, indirect band gaps in some of these materials, and the strain-induced transition in the gap types. Furthermore, it should be noted as ternary compounds, there could be many more 2D materials with this type of lattice structure. Indeed, in a recent work, more than 66 2D compounds were predicted to be stable with this structure, some of which are metals and are even magnetic~\cite{hong2020chemical,wang2020structure}. The analysis here may be extended to those predicted materials as well in future studies.

Besides Berry curvature, orbital magnetic moment $\bm m_{n\bm k} $ is another important geometric property of Bloch states~\cite{xiao2010berry}. For 2D systems, its direction is constrained to be out-of-plane, so like Berry curvature, it also behaves as a pseudoscalar. Using the effective model (\ref{kp2}), we can find that for the conduction bands
\begin{equation}
  m_c^\tau(\bm k, s_z)=-\tau\frac{2m_e\alpha^2 \tilde{\Delta}_{\tau s_z}}{{\hbar}^2\left(\tilde{\Delta}_{\tau s_z}^{2}+4 \alpha^{2} k^{2}\right)}\mu_B,
\end{equation}
where $\mu_B=e\hbar/(2m_e)$ is the Bohr magneton. Hence, the orbital moments are also opposite for the two valleys, consistent with the requirement of $T$ symmetry. Different from the Berry curvature, the moments for the valence bands are given by $m_v^\tau(\bm k, s_z)=m_c^\tau(\bm k, s_z)$, which do not have a flipped sign. A direct consequence of this moment is that under a magnetic field, the band structure will have an energy shift of $-\bm m\cdot\bm B$. Here, the shift will be opposite for the two valleys. This allows the control of valley polarization by a vertical magnetic field~\cite{cai2013magnetic}. Physically, this shift caused by orbital magnetic moment is closely related to the anomalous Landau levels, i.e., due to the opposite energy shift, the lowest Landau level for either conduction or valence band will have contribution from only one valley~\cite{cai2013magnetic}.

In conclusion, we have revealed interesting valley physics in the newly discovered 2D materials monolayer MoSi$_{2}$N$_{4}$, WSi$_{2}$N$_{4}$, and MoSi$_{2}$As$_{4}$. We show that these materials are semiconductors with an emerging binary valley degree of freedom.
By constructing the valley effective model, we demonstrate that these valleys are of Dirac-type, which therefore possess nontrivial valley-contrasting properties such as spin-valley coupling, Berry curvature, optical circular dichroism, and orbital magnetic moment.
These properties offer great opportunities to control valleys using applied electric, magnetic, and optical fields. Furthermore, we find that the band gap type and the valleys can be effectively switched by a moderate strain.
Our results thus reveal this new family of 2D materials as promising platforms for studying the valley physics and for applications in valleytronics and spintronics.

\begin{acknowledgements}
The authors thank X. D. Zhou and D. L. Deng for valuable discussions. This work is supported by the Singapore Ministry of Education AcRF Tier 2 (Grant No.~MOE2019-T2-1-001). S.G. acknowledges support from the NSF of China (Grants No.11904359). W.F. and Y.Y. acknowledge the support from the National Natural Science Foundation of China (Grants No. 11874085 and No. 11734003) and the National Key R\&D Program of China (Grant No. 2016YFA0300600). The computational work for this article was performed on resources of the National Supercomputing Centre, Singapore.
\end{acknowledgements}

\begin{appendix}

\section{Band structure results from hybrid functional approach}\label{A}
\begin{figure*}[htbp]
	\includegraphics[width=17.8cm]{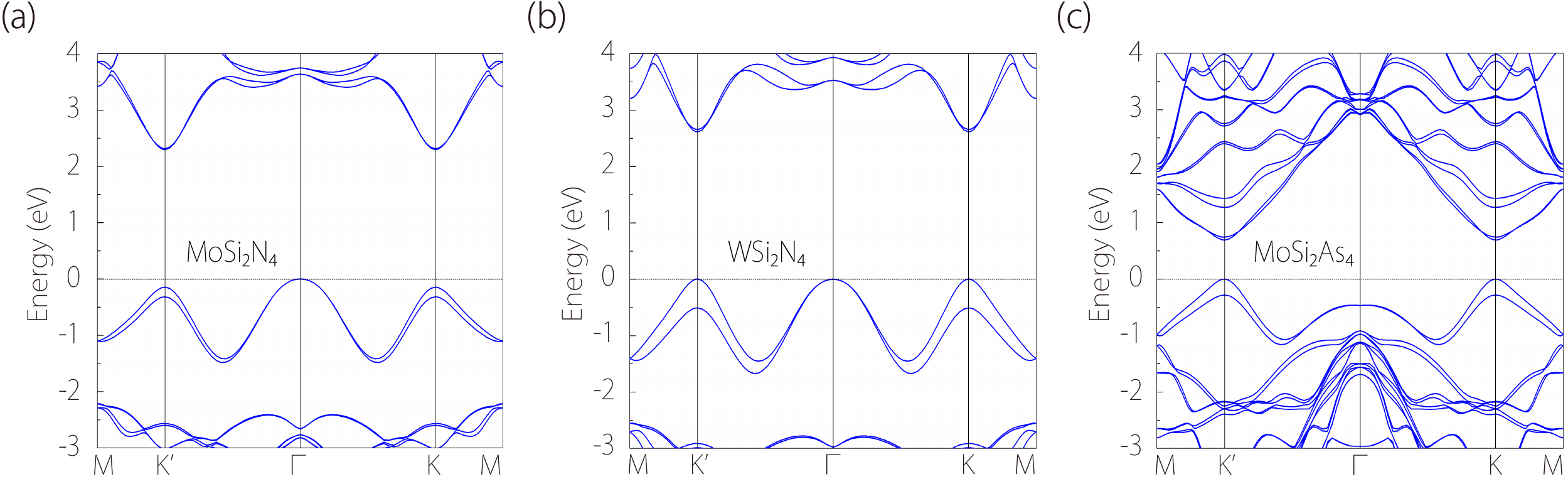}
	\caption{Band structures of monolayer (a) MoSi$_{2}$N$_{4}$, (b) WSi$_{2}$N$_{4}$, and (c) MoSi$_{2}$As$_{4}$ calculated with the hybrid functional approach (HSE06). SOC is included in the calculation.}
	\label{fig10}
\end{figure*}

The band structures of monolayer MoSi$_{2}$N$_{4}$, WSi$_{2}$N$_{4}$ and MoSi$_{2}$As$_{4}$ are also checked by using the more sophisticated Heyd-Scuseria-Ernzerhof hybrid functional approach (HSE06)~\cite{heyd2003hybrid} (SOC included). The results are plotted in Fig.~\ref{fig10}. One observes that qualitative band features are similar to the PBE results (especially for the low-energy bands at the $K$/$K'$ valleys), except that for WSi$_2$N$_4$, the valence band at $K$/$K'$ now becomes slightly higher than $\Gamma$.

\end{appendix}

%\bibliography{MoSi2N4_ref}


\end{document}